# On The Critical Reynolds Number For Transition From Laminar To Turbulent Flow

Trinh, Khanh Tuoc

Institute of Food Nutrition and Human Health

Massey University, New Zealand

*K.T.Trinh@massey.ac.nz*

## Abstract

In this visualisation, the transition from laminar to turbulent flow is characterised by the intermittent ejection of wall fluid into the outer stream. The normalised thickness of the viscous flow layer reaches an asymptotic value but the physical thickness drops exponentially after transition. The critical transition pipe Reynolds number can be obtained simply by equating it with the asymptotic value of the normalised thickness of viscous flow layer.

Key words: Transition, critical stability Reynolds number, critical transition Reynolds number, non-Newtonian pipe flow

# 1 Introduction

Reynolds (1883) was the first to propose a criterion for differentiation between laminar and turbulent flows in his classic dye visualisation

$$\text{Re} = \frac{DV\rho}{\mu} \tag{1}$$

and suggested a critical value of $\text{Re} = 2100$ for the upper limit of laminar flow. In a second paper (Reynolds, 1895) he showed by time-averaging the Navier-Stokes equations that new extra convection terms appeared in turbulence which have the units of stress and are therefore called Reynolds stresses. Considerable effort has been expanded in the last hundred and thirty years to understand the process of transition. "The problem is simple in concept and yet the origins of the observed turbulent motion remain largely mysterious despite more than a century of research" (Mullin and Peixinho, 2006). The literature on the transition is vast, prompting Herbert (1988), a well-known author, to note that "different reviews on shear-flow instability may have little in common and a zero-overlap of cited literature. This curious fact illustrates the many facets of the overall problem, the multitude of views, concepts, and methods, and the need to remain open minded. It also grants me the right to present my own view supported by a selection of references that I know is far from complete". We will therefore not even attempt a survey of the present literature but will concentrate on some developments directly relevant to the topic of this paper.

Kerswell (2005) noted that all experimental and theoretical evidence points to the fact that the laminar flow state (which exists potentially for all flow rates) is linearly stable to any infinitesimal disturbance. He also stated that the clear implication is that the observed transition process can then only be initiated by finite amplitude disturbances. This view is widely shared by other authors.

Reynolds himself observed that turbulence was triggered by inlet disturbances to the pipe and the laminar state could be maintained to Re ≈12,000 if he took great care in minimizing external disturbances to the flow. By careful design of pipe entrances Ekman (1910) has maintained laminar pipe flow up to a Reynolds number of 40,000 and Pfenniger (1961) up to 100,000 by minimising ambient disturbances. All sufficiently small perturbations will decay e.g. Salwen et al. (1980) and Meseguer & Trefethen (2003), who analyzed the problem up to $Re = 10^7$. Turbulence is only triggered when the disturbances exceeded a finite critical amplitude of perturbation Mullin and Peixinho (2006). Thus, to trigger transition, two thresholds have to be crossed: The flow has to be sufficiently fast and a perturbation has to be strong enough. Eckhardt et al. (2007) further noted that experiments by Hof et al. (2003) further show that as the Reynolds number increases the critical threshold decreases so that at sufficiently high Reynolds numbers the unavoidable residual fluctuations always suffice to trigger turbulent flow.

Bayly et al.(1988) have reviewed very well the development of theoretical thinking on the stability of shear flows. Its basis can be traced to Reynolds'concept that the laminar pattern always represents a possible type of flow since it is a solution of the Navier-Stokes equation but can be broken into a turbulent pattern by growing perturbations.

The fundamental differential equation for the disturbance or stability equation was extracted from the Navier-Stokes equations by Orr (1907) and Sommerfield (1908)

$$(U - c)(\phi'' - \alpha^2 \phi) - U''\phi = \frac{i}{\alpha \mathrm{Re}}(\phi'''' - 2\alpha^2 \phi'' + \alpha^4 \phi) \qquad (2)$$

where the disturbance has the form

$$\psi(x, y, t) = \phi(y) e^{i(\alpha x - \beta t)} \qquad (3)$$

$\alpha$ is a real number and $\lambda = 2\pi/\alpha$ is the wavelength of the disturbance. The quantity $\beta$ is complex

$$\beta = \beta_r + i\beta_i \qquad (4)$$

where $\beta_r$ is the circular frequency of the partial disturbance and $\beta_i$ is the amplification factor. The results are often expressed in terms of the ratio of $\alpha$ and $\beta$

$$c = \frac{\beta}{\alpha} = c_r + ic_i \qquad (5)$$

Over a century ago, Lord Rayleigh (1880) analysed the problem for inviscid flow where the right hand side of equation (2) can be neglected and formulated two important theorems

1. Velocity profiles which posses an inflexion point are unstable

2. The velocity of propagation of neutral disturbances $(c_i = 0)$ in a boundary layer is smaller than the maximum velocity of the main flow $U_m$

In the last hundred years much effort has been directed to solving the Orr-Sommerfeld equation for various flows but this equation is exceedingly difficult to analyze for the large Reynolds numbers at which transition is observed to occur. Thus not much progress were made until Tollmien and Schlichting obtained predictions for a curve of neutral stability (Schlichting, 1979) that was verified experimentally a decade later by Schubauer and Skramstad (1943) as shown in Figure 1.

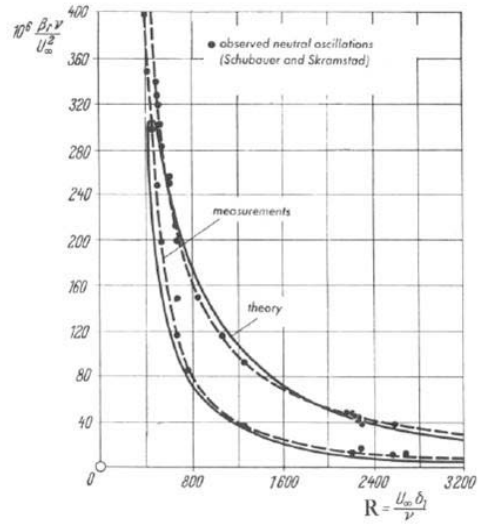

Figure 1 Curve of neutral stability for neutral disturbances on a flat plate at zero incidence from Schubauer and Skramstad (1943).

The minimum Reynolds number on the stability curve on flat plate is

$$\text{Re}_{\delta_1,s} = \left(\frac{U_\infty \delta_1}{\nu}\right)_{cs} = 520 \qquad (6)$$

where $\delta_1$ is the displacement thickness. At lower Reynolds numbers all disturbances are dissipated but turbulence does not set in at that point. From this critical stability Reynolds number onwards the disturbance can grow but transition to turbulence occurs at $(U_\infty x/\nu) = 3.5 \times 10^5 - 10^6$ corresponding to

$$\text{Re}_{\delta_1,T} = 950 \qquad (7)$$

For plane Poiseuille flow Lin (1955) predicted a critical stability Reynolds number of 5314. Thus the linear instability theory gives a satisfactory description of the initial growth of very small disturbances but does not predict well the Reynolds numbers at which transition is observed. In the linear instability theory the most dangerous instabilities have fairly long wavelengths and low growth rates, in contrast to the universally observed three-dimensional short-wavelength structure of turbulence. The discrepancy may be caused by the neglect of the nonlinear self-interaction of the disturbance, which always exists for finite perturbations no matter how small. The next advance in the theory of stability was the modelling of non-linear two dimensional waves but from the time of Reynolds' earliest experiments, it was obvious that three dimensional disturbances are important at some stage in transition. Three dimensional disturbances have been studied extensively since the 1960's as they provide a crucial link between classical instability theory and the observations of transition in real flows. These developments have been well reviewed by Bayley et al. (1988).

Eckhardt et al. (2007) have reviewed another approach based on the theory of strange attractors. The background for these studies is the abstraction to consider the system in its state space. In state space, there is one region dominated by the laminar flow. The time-independent parabolic profile is a fixed point in this space. The parabolic profile is linearly stable and, hence, all points in its neighbourhood evolve toward this fixed point; these states form the basin of attraction of the laminar flow. The turbulent dynamics take place in other parts of the state space where a turbulence basin of attraction is situated. The spatially and temporally fluctuating dynamics of the turbulent regions suggests that there are chaotic elements, such as horseshoe vortices, just as in a regular attractor. However, the possibility of decay indicates that the basin is not compact nor space filling; there must be connections to the laminar profile. In dynamical systems such structures are known as chaotic saddles or strange saddles (Eckhart et al., op. cit.).

Kerswell (2005) has reviewed another approach based on the developments and promise of the recently discovered travelling wave solutions. Mullin and Peixinho (2006) have reviewed the results of recent experimental investigations into transition to turbulence in fluid flow through a circular straight pipe, at room temperature.

The failure of the formal approach to predict accurately the well known critical Reynolds number for transition from laminar to real turbulent flows in different geometries have led some authors to take a more heuristic approach that allows an approximate estimate of the critical transition Reynolds number in situations of practical interest. Of particular interest is the transition in non-Newtonian fluids for which experimental data is still scarce and inaccurate e.g. Ryan and Johnson (1963a), Hanks (1963a), Mishra and Tripathi (1971, 1973), Rudman et al.(2002) for power law fluids, Hanks (1963a) for Bingham plastic fluids, Zamora et al (2005), Peixinho et al (2005) for Herschel-Bulkley fluids and other geometries such as square ducts (Etemad and Sadeghi, 2001). Wilson and Thomas (2006) tried to define an equivalent viscosity for Bingham plastic fluids to better identify the transition Reynolds number, Güzel et al. (2009) defined a new Reynolds number based on averaging local Reynolds numbers across the pipe. Ryan and Johnson proposed a stability criterion

$$Z = \frac{R\rho U}{\tau_w} \frac{\partial U}{\partial y} \qquad (8)$$

which is essentially an empirical estimate of the ratio of rate of energy supply to the disturbance to rate of dissipation. $Z$ must clearly be zero at the wall and the centreline and passes through a maximum at an intermediate position. For a flow where the laminar velocity profile is known $Z_{max}$ can be calculated. For Newtonian pipe flow this coefficient can be shown to be a special case of the Reynolds number

$$Z_{max} = \sqrt{\frac{4}{27}} \frac{DV}{\nu} \quad at \quad \frac{r}{R} = \frac{1}{\sqrt{3}} \qquad (9)$$

The critical stability parameter for pipe flow is given at $Re_c = 2100$ as

$$Z_c = 808 \qquad (10)$$

Ryan and Johnson assumed that this value of $Z_c$ also applied to non-Newtonian fluids. Hanks and his co-workers have applied the analysis of Ryan and Johnson to power law fluids in concentric annuli and parallel plates (Hanks, 1963b, Hanks and Valia, 1971, Hanks and Ricks, 1975) and Bingham plastic fluids (Hanks, 1963a, Hanks and Dadia, 1971). For power law fluids Hanks predicted a critical Reynolds number of

$$Re_{g,c} = 6464 \frac{n(2+n)^{((2+n)/(1+n))}}{(1+3n)^2} \qquad (11)$$

where $Re_g$ is the generalised Metzner-Reed Reynolds number (Metzner and Reed, 1955). Mishra and Tripathi (1971, 1973) observed that the predictions of Ryan and Johnson and Hanks were acceptable for moderate values of the behaviour index $n$ of power law fluids but did not agree with experimental evidence for highly non-Newtonian fluids (low $n$ values) as shown in Figure 15. They proposed a different empirical criterion based on the ratio of average kinetic energy of the fluid and friction drag at the wall

$$\alpha_c = \frac{\rho V^2}{Average\ K.E.\ per\ unit\ volume} \tag{12}$$

$$Re_{g,c} = 2100\alpha_c \tag{13}$$

$$f_c = \frac{16}{Re_{g,c}} = \frac{0.0762}{\alpha_c} \tag{14}$$

For power law fluids

$$Re_{g,c} = 2100\frac{(4n+2)(5n+3)}{3(3n+1)^2} \tag{15}$$

## 2  Theoretical considerations

Following the advice of Herbert, and with the indulgence of the reader, I will present the following views from a personal perspective and not use the traditional third person narrative because my views have clear points of difference with the traditional approach.

### 2.1  The four component decomposition of the instantaneous velocity in turbulent flows and its implications

In previous publications (Trinh, 1992, 2009c, 2009a) I have proposed that much more information can be extracted from measurements of turbulent flows by decomposing the instantaneous velocity into four components instead of Reynolds' two components. The wall layer process has been well documented by many authors using hydrogen bubble tracers (Kline et al., 1967, Kim et al., 1971, Corino and Brodkey, 1969) as shown schematically in Figure 2. It starts with an inrush of fast fluid towards the wall that is deflected into a longitudinal vortex. The fluid beneath the vortex is represented by a low-speed streak that develops during the so-called sweep phase. The streaks tend to lift, oscillate and eventually burst in other violent ejections from the wall towards the outer region. The sweep phase is much longer lasting than the burst phase.

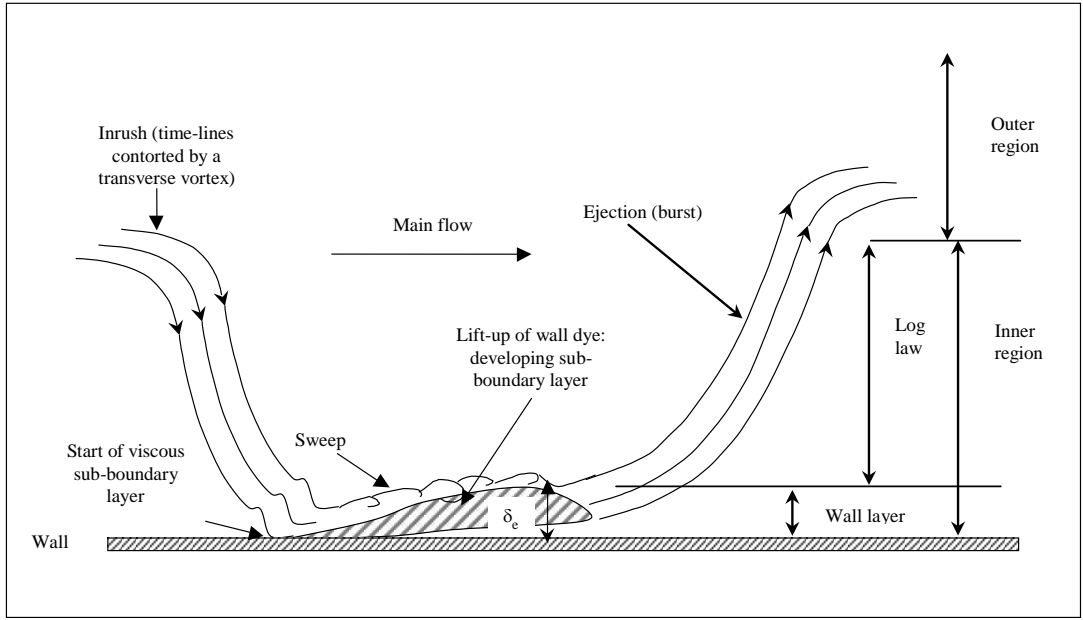

Figure 2. Visualisation of a cycle of the wall layer process drawn after the observations of Kline et al.(1967) and regions in the flow field.

Figure 3 shows a trace of the instantaneous velocity $u_i$ corresponding to a cycle in the wall layer. The local instantaneous velocity $u_i$ can be decomposed into a smoothed velocity $\tilde{u}_i$ in the sweep phase of the wall layer and a fast fluctuating velocity $u'_i$ resulting from the passage of the longitudinal vortex above the wall.

$$u_i = \tilde{u}_i + u'_i \tag{16}$$

where $u'_i$ is periodic and does not contribute to the long time average velocity $U_i$

$$\int_0^\infty u'_i dt = 0 \tag{17}$$

$$U_i = \frac{1}{t_\nu} \int_0^{t_\nu} \tilde{u}_i \, dt \tag{18}$$

$$u_i = U_i + U'_i \tag{19}$$

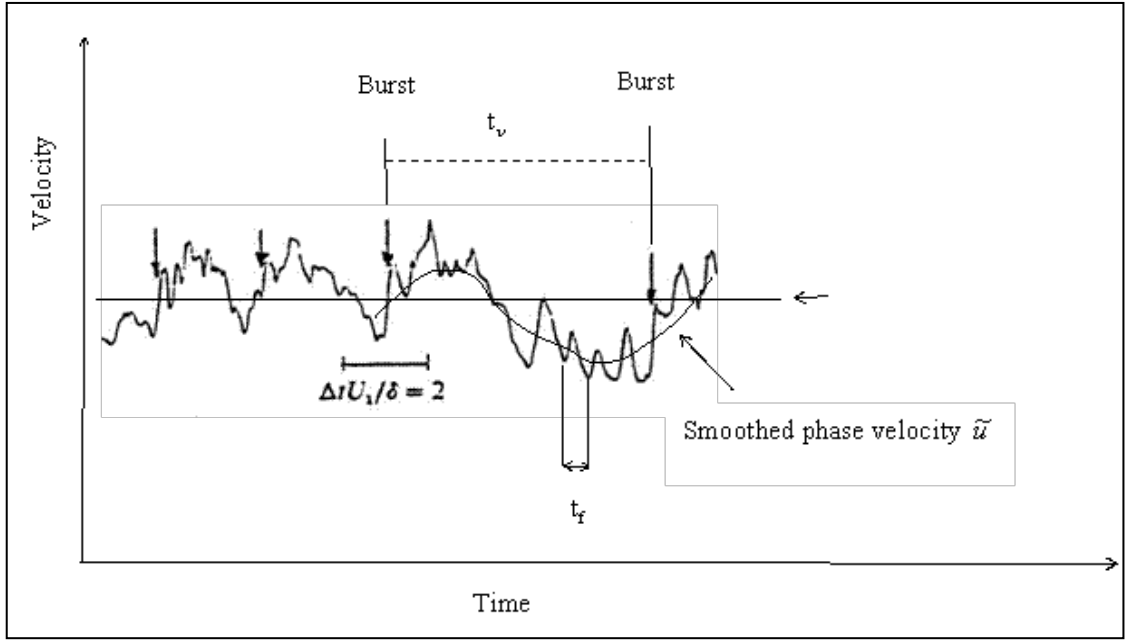

Figure 3 Trace of instantaneous streamwise velocity in the wall layer after measurements by Antonia et al. (1990).

$$U'_i = \tilde{U}'_i + u'_i \tag{20}$$

where

$$\tilde{U}'_i = \tilde{u}_i - U_i \tag{21}$$

then

$$u_i = U_i + \tilde{U}'_i + u'_i \tag{22}$$

We may average the Navier-Stokes equations over the period $t_f$ of the fast fluctuations. Bird, Stewart, & Lightfoot (1960), p. 158 give the results as

$$\frac{\partial(\rho \tilde{u}_i)}{\partial t} = -\frac{\partial p}{\partial x_i} + \mu \frac{\partial^2 \tilde{u}_i}{\partial x_j^2} - \frac{\partial \tilde{u}_i \tilde{u}_j}{\partial x_j} - \frac{\partial \overline{u'_i u'_j}}{\partial x_j} \tag{23}$$

Equation (23) defines a second set of Reynolds stresses $\overline{u'_i u'_j}$ which we will call "fast" Reynolds stresses to differentiate them from the standard Reynolds stresses $\overline{U'_i U'_j}$.

We may write the fast fluctuations in the form

$$u'_i = u_{0,i}\left(e^{i\omega t} + e^{-i\omega t}\right) \tag{24}$$

The fast Reynolds stresses $u'_i u'_j$ become

$$u'_i u'_j = u_{0,i} u_{0,j}(e^{2i\omega t} + e^{-2i\omega t}) + 2 u_{0,i} u_{0,j} \tag{25}$$

Equation (25) shows that the fluctuating periodic motion $u'_i$ generates two components of the "fast" Reynolds stresses: one is oscillating and cancels out upon long-time-averaging, the other, $u_{0,i} u_{0,j}$ is persistent in the sense that it does not depend on the period $t_f$. The term $u_{0,i} u_{0,j}$ indicates the startling possibility that a purely oscillating motion can generate a steady motion which is not aligned in the direction of the oscillations. The qualification steady must be understood as independent of the frequency ω of the fast fluctuations. If the flow is averaged over a longer time than the period $t_v$ of the bursting process, the term $u_{0,i} u_{0,j}$ must be understood as transient but non-oscillating. This term indicates the presence of transient shear layers embedded in turbulent flow fields and not aligned in the stream wise direction similar to those associated with the streaming flow in oscillating laminar boundary layers (Schneck and Walburn, 1976, Tetlionis, 1981). Thus a fourth component must be included in the decomposition of the instantaneous velocity in turbulent flows

$$u_i = U_i + \tilde{U}'_i + u'_{i}(\omega t) + u_{i,st} \qquad (26)$$

The streaming flow term $u_{i,st}$ dominates the bursting phase.

## 2.2 Evolution from two to four component velocities

To understand the transition process we must, in my view, show how the solution of the Navier-Stokes transforms from the laminar state characterised by a two-component instantaneous velocity (equation 16) to the fully turbulent state characterised by a four-component instantaneous velocity (equation 26). This evolution also occurs in the process of turbulent wall shear flow which we consider first. The analytical solution is extremely difficult and much greater progress has been made by direct numerical simulations DNS. Unfortunately the database generated by the DNS is usually massive and we cannot easily trace the evolution of the terms in equation (26) let alone their interactions. Robinson (1991) for example spent an entire PhD simply to describe and catalogue the coherent structures that identified from the DNS of Moin and Kim (1982).

The sweep phase can be modelled as a Kolmogorov flow (Obukhov, 1983) a simple two dimensional sinusoidal flow, or better still analysed with techniques borrowed from laminar oscillating flow (Trinh, 2009c, Trinh, 1992). The traditional approach to analyse these unsteady flows is by a method of successive approximations (Schlichting, 1979, Tetlionis, 1981). The dimensionless parameter defining these successive approximations is

$$\varepsilon = \frac{U_e}{L\omega} \qquad (27)$$

where $U_e$ is the local mainstream velocity and L is a characteristic dimension of the body. The smoothed velocity $\tilde{u}_i$ is given by the solution of order $\varepsilon^0$ which applies when $\varepsilon \ll 1$. The governing equation (Einstein and Li, 1956, Hanratty, 1956, Meek and Baer, 1970, Trinh, 2009c) is a subset of the NS equations

$$\frac{\partial \tilde{u}}{\partial t} = \nu \frac{\partial^2 \tilde{u}}{\partial y^2} \qquad (28)$$

where $\tilde{u}$ refers here to the smoothed velocity $\tilde{u}_i$ in the $x$ direction. It does not require that there are no velocity fluctuations, only that they are small enough for their effect on the smoothed phase velocity $\tilde{u}$ to be negligible. Stokes (1851) has given the solution to equation (28) as

$$\frac{\tilde{u}}{U_\nu} = \operatorname{erf}(\eta_s) \qquad (29)$$

where $\eta_s = \frac{y}{\sqrt{4\nu t}}$. The thickness of this sub-boundary layer is

$$\delta_\nu^+ = 4.16 U_\nu^+ \qquad (30)$$

where the velocity and normal distance have been normalised with the wall parameters $\nu$ the kinematic viscosity and $u_* = \tau_w/\rho$ the friction velocity, $\tau_w$ the time averaged wall shear stress and $\rho$ the density.

To understand the effect of the fluctuations beyond simple growth, an analysis of oscillating flow with a zero-mean velocity is particularly interesting since the basic velocity fluctuations imposed by external means do not grow with time because there is no mean motion along the wall. We may thus investigate the effect of the amplitude and frequency of the fluctuations separately. The following treatment of the problem is taken from the excellent book of (Tetlionis, 1981).

We define a stream function $\psi$ such that

$$u = \frac{\partial \psi}{\partial y} \qquad v = \frac{\partial \psi}{\partial x} \qquad (31)$$

The basic variables are made non-dimensional

$$x^* = \frac{x}{L} \qquad y^* = \frac{y}{\sqrt{2\nu/\omega}} \qquad t^* = t\omega \qquad (32)$$

$$U_e^*(x,t) = \frac{U_e}{U_\infty}(x,t) \qquad \psi^* = \psi \left( U_\infty \sqrt{\frac{2\nu}{\omega}} \right)^{-1} \qquad (33)$$

where $U_\infty$ is the approach velocity for $x \to \infty$. The system of coordinates x, y is attached to the body. The Navier-Stokes equation may be transformed as:

$$\frac{\partial^2 \psi^*}{\partial y^* \partial t^*} - \frac{1}{2}\frac{\partial^3 \psi^*}{\partial y^{*3}} - \frac{\partial U_e^*}{\partial t^*} = \frac{U_e}{L\omega}\left( -\frac{\partial \psi^*}{\partial y^*}\frac{\partial^2 \psi^*}{\partial y^* \partial x^*} + \frac{\partial \psi^*}{\partial x^*}\frac{\partial^2 \psi^*}{\partial y^{*2}} + U_e^* \frac{\partial U_e^*}{\partial x^*} \right) \qquad (34)$$

with boundary conditions

$$\psi^* = \frac{\partial \psi^*}{\partial y^*} = 0 \qquad y^* = 0 \qquad (35)$$

For large values of $\omega$, the RHS of equation (34) can be neglected since

$$\varepsilon = \frac{U_e}{L\omega} \ll 1 \qquad (36)$$

In this case, Tetlionis reports the solution of equation (34) as:

$$\psi^* = \left[ \frac{U_0^*(x^*)}{2}(1-i)[1 - e^{(1+i)y^*}] + \frac{U_0^* y^*}{2} \right] e^{it^*} + C \qquad (37)$$

Tetlionis (op. cit. p. 157) points out that equation (37) may be regarded as a generalisation of Stokes' solution (1851) for an oscillating flat plate often called Stokes second problem. This latter solution describes an oscillating flow called the Stokes layer which is often found embedded in other flow fields and has properties almost independent of the host field.

In the case of a non-zero mean, the smoothed velocity can be described by a generalised form of Stokes solution for a flat plate started impulsively, often referred to as Stokes' first problem (equation 28). Then the stream function must incorporate both of Stokes' solutions. This composite solution, that we called of solution order $\varepsilon^0$, is accurate only to an error of order $\varepsilon$. Tetlionis reports a more accurate solution for the case when $\varepsilon$ cannot be neglected (i.e. for larger $\varepsilon$ hence lower frequencies):

$$\psi^* = \frac{U_0^*(x^*)}{2}[\psi_0^*(y^*)e^{it^*} \overline{\psi_0^*}(y^*)e^{-it^*}] + \varepsilon[\psi_1^*(x^*,y^*)e^{2it^*} + \overline{\psi_0^*}e^{-2it^*}] + O(\varepsilon^2) \quad (38)$$

where $\psi_0$ and $\psi_1$ are the components of the stream function of order $\varepsilon^0$ and $\varepsilon$. Substituting this more accurate solution into equation (34), we find that the multiplication of coefficients of $e^{it^*}$ and $e^{-it^*}$ forms terms that are independent of the oscillating frequency, ω, imposed on the flow field and were not anticipated in equation (38). Thus the full solution of equation (34) is normally written (Tetlionis, 1981) as

$$\psi^* = \frac{U_0^*(x^*)}{2}[\psi_0^*(y^*)e^{it^*} + \overline{\psi_0^*}(y^*)e^{-it^*}]$$
$$+ \varepsilon[\psi_{st}^* + [\psi_1^*(x^*,y^*)e^{2it^*} + \overline{\psi_1^*}(x^*,y^*)e^{-2it^*}] + O(\varepsilon^2) \quad (39)$$

where the overbar denotes the complex conjugate and $\psi_{st}^*$ results from cancelling of $e^{it^*}$ and $e^{-it^*}$ terms.

The quantity $\psi_{st}^*$ shows that the interaction of convected inertial effects of forced oscillations with viscous effects near a wall results in a non-oscillating motion that is referred to in the literature as "Streaming". The problem has been known for over a century (Faraday, 1831, Dvorak, 1874, Rayleigh, 1884, Carriere, 1929, Andrade, 1931, Schlichting, 1932) and studied theoretically (Riley, 1967, Schlichting, 1960, Stuart, 1966, Tetlionis, 1981). The existence of this streaming flow, even in this absence of any mainstream flow, is clearly demonstrated in Figure 4.

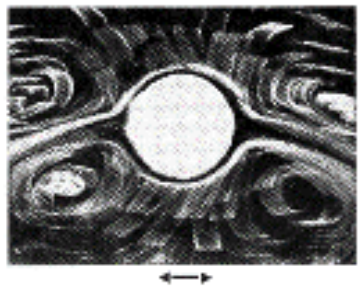

Figure 4 Streaming flow near a vibrating cylinder. After Schlichting (1960).

At the end of the sweep phase, the fluctuations have grown large enough for the streaming flow to contain substantial amount of kinetic energy sufficient to eject wall fluid from the wall layer.

The analytical solution of the stream function $\psi_{st}^*$ e.g. by Stuart (1966) presents great difficulties and invariably involve simplifications that suppress critical information. For example Stuart (1966) ignored the interaction terms between the main flow and the streaming flow. The main crossflow deflects the wall ejection in a streamwise direction. This is the first interaction. Immediately outside the wall layer, the jet follows a quasi linear path before it becomes deflected into a curvilinear pattern. As the main flow hits the jet, it is broken into small scales typical of incoherent turbulence. By contrast the jet itself moves as a coherent structure.

**2.3 Physical visualisation of transition**

The ejections start to disturb the outer quasi-inviscid region beyond the wall layer and dramatically increase the boundary layer thickness. At Reynolds numbers just above the critical value, e.g. Re =2100 for pipe flow, only the far field section of the intermittent jets penetrates the outer region. The disturbance to the previously "quasi-potential" flow may be compared with that of a wall-parallel jet since the ejections are here aligned in the direction of main flow. This region has been described by Cole's law of the wake (Coles, 1956). As the Reynolds number increases further, so do the fluctuations: the streaming flow strengthens and emerges at a cross flow angle with the main flow. Millikan (1938) showed that an outer region that scale with the outer parameters (Coles law of the wake in the present visualisation) and a wall region which scales with the wall parameters (the solution of order $\varepsilon^0$) must be linked by a semi-logarithmic velocity profile.

$$U^+ = A \ln y^+ + B = \frac{1}{\kappa} \ln y^+ + B \qquad (40)$$

where $\kappa$ is called Karman's universal constant with a canonical value of *0.40*. In the present visualisation, upon transition, the first layer to be added to the wall layer is the law-of-the-wake region then full turbulence is established when the log-law grows.

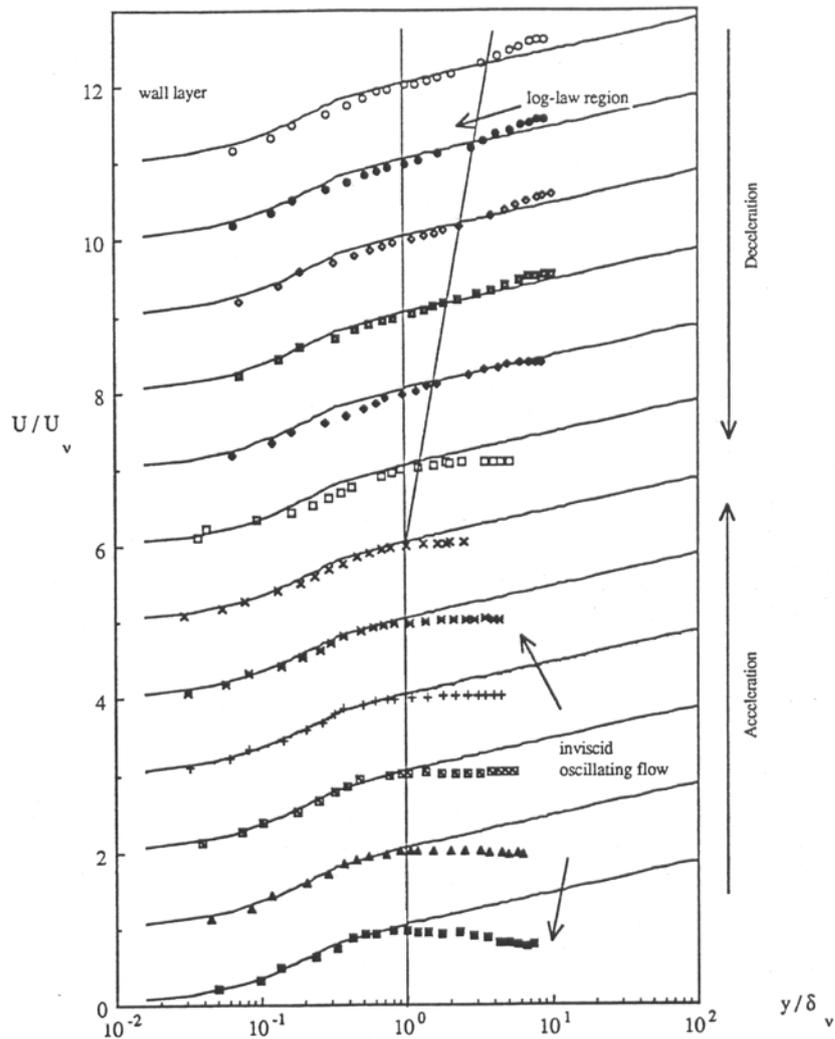

Figure 5. Penetration of the log-law into the outer region during a cycle oscillating pipe flow. From Trinh (1992). Data of Akhavan et al. (1991) for $Re_\omega = 1080$.

The transition from laminar to turbulent flow is clearly seen in the oscillating flow experiments of Akhavan, Kamm, & Saphiro Akhavan (1991) . The flow is driven by a reciprocating piston pump. The acceleration phase, where the pressure gradient is favourable, is laminar. The velocity profile here exhibits only two regions: (a) a wall layer which coincides very well with the profiles for laminar boundary layer flow and those for the wall layer of steady turbulent pipe flow, and (b) a fluctuating potential flow in the outer region.

Schneck and Walburn (1976) have argued in their study of pulsatile blood flow that the secondary streaming flow results from a tendency of viscous forces to resist the reversal of flow imposed by the oscillating motion of the main stream. This is demonstrated more clearly in the experiments of Gad-el-Hak, Davis, McMurray, & Orszag (1983) who generated an artificial bursting process in a laminar boundary layer on a flat plate by decelerating it.

The magnitude of the deceleration oscillating flow and the corresponding adverse pressure gradient must be sufficient to induce separation and ejection of low-speed fluid from the wall. The ejections begin to penetrate the previously quasi-inviscid region outside but because their energy content is still weak they are immediately deflected in the direction of the main flow: The law of the wake layer makes it appearance. The transition region ends when the ejections have gathered enough strength for the rectilinear region of the jet to protrude from the wall layer: the log-law layer appears. The growth of the log-law region in between the wall layer and the law of the wake during the decelerating phase of oscillating flow is seen clearly in Figure 5 where the original data of Akhavan et al. has been rearranged.

The evolution of the three layers (wall buffer layer, log-law layers and Cole's law of the wake layer) on a flat plate with increasing Reynolds number can be calculated from published velocity profiles using one of several methods that are mutually compatible (Trinh, 2010a) as shown in Figure 7.

## 2.4 Characteristics of the present view on transition

Since the seminal paper of Reynolds (1883), many researchers have approached transition as "the issue of how and why the fluid flow along a circular pipe *changes* from being laminar (highly ordered in space and time) to turbulent (highly disordered in both space and time) as the flow rate increases" (Kerswell 2005). This view that turbulence breaks down the previously laminar flow field is widely held. For example Bayley et al. (1988) state that "The process by which turbulent flow develops and *replaces* laminar flow is known as (turbulent) transition".

The analysis in section 2.2 suggests however that the solution of order $\varepsilon^0$ does not disappear from the flow field even as it evolves from a two component to a four component local instantaneous velocity. If we define laminar flow as one that only involve exchange of viscous momentum between the wall and the fluid, which would include unsteady oscillating laminar flow observed for example by Akhavan et al. (1991), rather than focus narrowly on steady state laminar flows described by the Hagen Poiseuille solution for pipes or Blasius solution for flat plates, then the persistence of the solution order $\varepsilon^0$ is fully compatible with the well known observation that the laminar flow state (which exists for all flow rates) is linearly stable to any infinitesimal disturbance. The inevitable conclusion in my mind was that *the development of turbulence does not break down or replace the viscous (laminar) layer represented by the solution of order $\varepsilon^0$ but actually transforms the previously quasi inviscid flow outside that layer.* This is a startling departure from the approach pioneered by Reynolds. His experiment and others appear to indicate that indeed the laminar flow field is completely disrupted and replaced at higher Reynolds numbers by a turbulent apparently chaotic flow field. For example, Mullin and Peixinho (2006) give a clear illustration of the effect of a turbulent puff on the flow field (Figure 6).

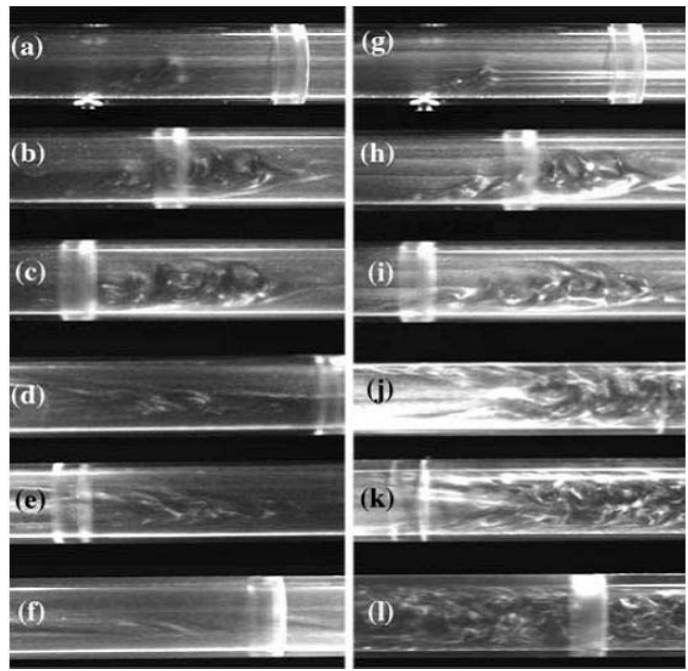

Figure 6. Development of a perturbation where a small amount of fluid was impulsively injected into the bulk flow through a 3mm hole. Re = 2000 and in the snapshot sequence (a)–(f) the perturbation decays. Images taken at 1, 2, 3, 6, 14 and 30 diameters downstream from the injection point using a camera travelling at the mean speed of the flow. In the sequence (g)–(l) a slightly larger amount of fluid was injected and a puff is created.

How does one reconcile these two views? The first clues to this perplexing issue came when I started to estimate the thicknesses of the different layers in various turbulent flows in 1986-7 and eventually compiled Figure 7 reproduced from (Trinh, 2010a). The reader should note that the normalised thickness of the buffer layer (the time-averaged value of the wall layer thickness) grows up to the critical transition Reynolds number then remained constant (Figure 7c). Many authors have shown that the velocity profile in this layer is well described by the Stokes solution in equation (29) (Einstein and Li, 1956, Meek and Baer, 1970, Trinh, 2005, Trinh, 2009a).

However since the normalised thickness of the boundary layer grows significantly upon transition the solution of order $\varepsilon^0$ becomes a smaller proportion of the entire flow field. Since the wall shear stress $\tau_w$ and the shear velocity $u_* = \sqrt{\tau_w/\rho}$ increase rapidly after transition, a constant normalised wall or buffer layer results in a fast decreasing physical thickness for this solution of order $\varepsilon^0$. In pipe flow for example, this solution occupies the entire pipe up to $Re = 2100$ then diminishes in physical thickness dramatically with Reynolds number as shown for pipe flow in Figure 8. Thus an observer or probe situated at the location Ob in Figure 8 may feel as though the flow at that location has broken down to smaller turbulence scales but in fact the solution of order $\varepsilon^0$ has not been altered; it has simply moved out of the field of vision of the observer.

The shape of the interface between the wall layer and the log-law layer in the vicinity of the critical transition Reynolds number is quite similar to the shape factor for a flat plate boundary layer as shown in Figure 8.

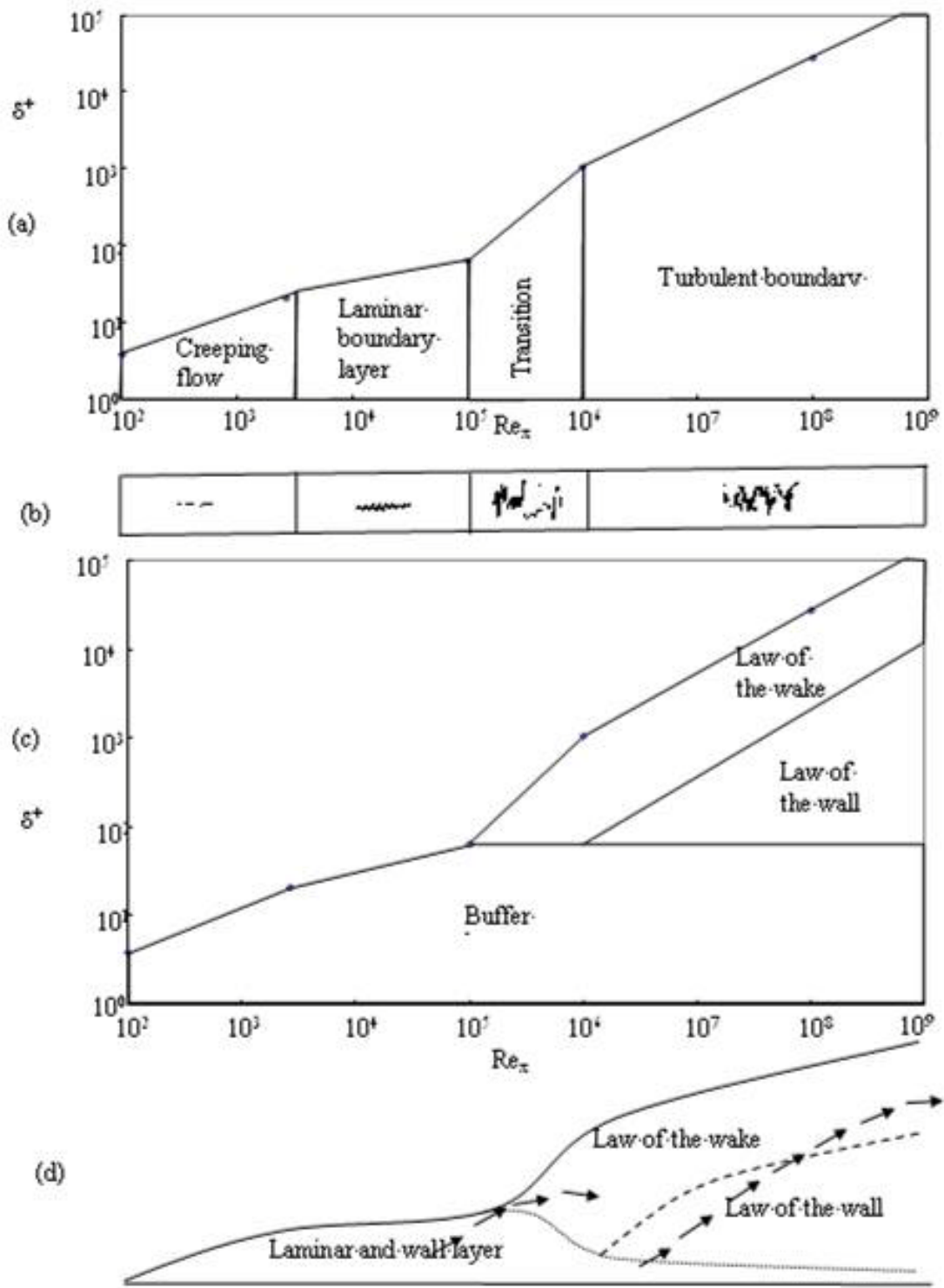

Figure 7 Representations of flow regimes (a) Reynolds flow regimes, (b) Representative velocity traces (after Schubauer and Skramstadt (1947), (c) Additive layers in normalised dimensions, (d) Physical thickness of layers (not to scale) and path of ejections in transition and fully turbulent flow.

There is however a fundamental difference between the Hagen Poiseuille and Blasius laminar flows which are steady state solutions of order $\varepsilon^0$ for $t_\nu \to \infty$ and the wall layer flow which is an unsteady state solution of order $\varepsilon^0$ with a much shorter time scale. The streaming flow expels fluid under the low-speed streaks when their kinetic energy has reached a critical level.

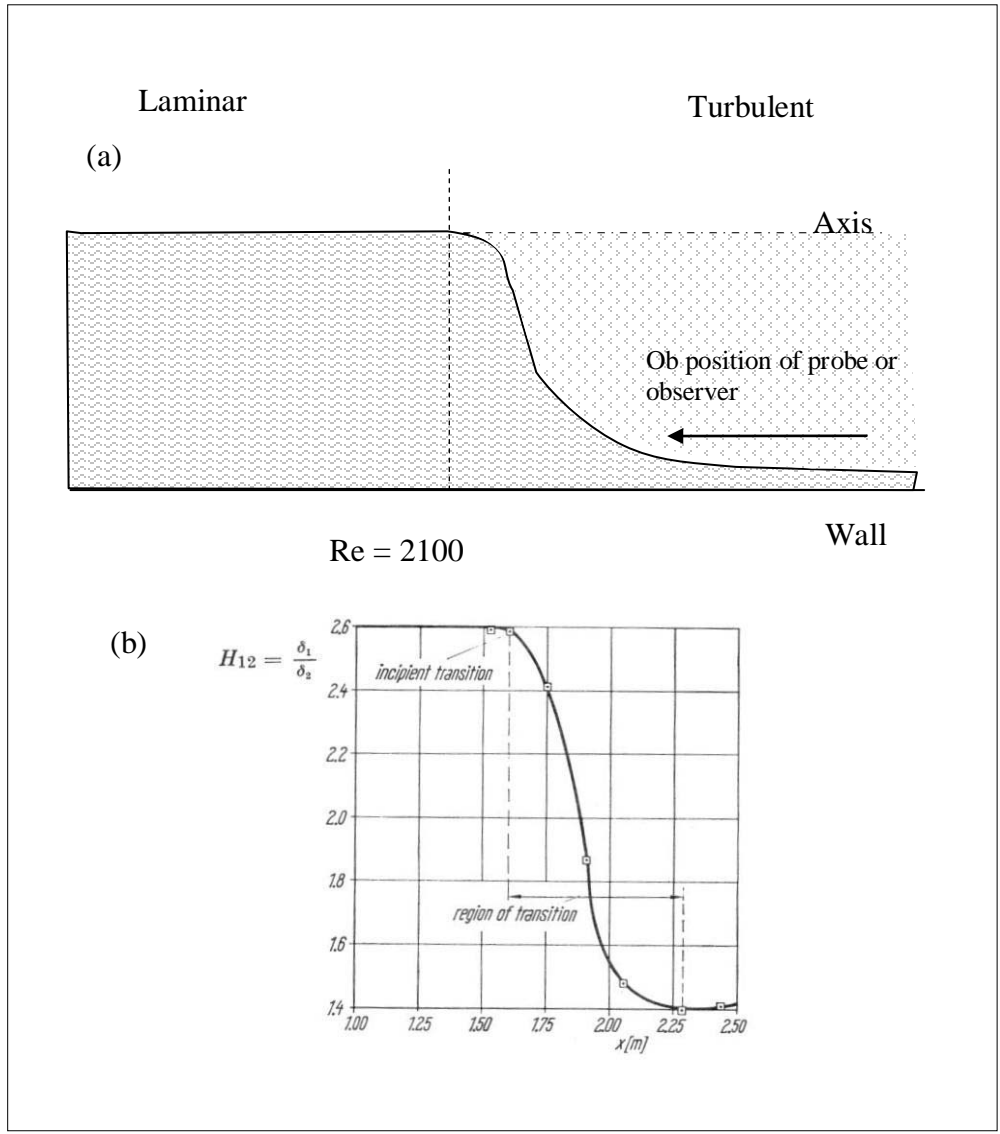

Figure 8 (a) Variation of physical value of wall layer thickness in pipe flow with Reynolds number (not to scale) and (b) boundary layer shape factor measured by Schubauer and Klebanoff as quoted by Schlichting (1960).

This disrupts the solution of order $\varepsilon^0$ and triggers an inrush of fluid from the mainstream to start a new sweep phase. In that sense, the flow field is periodically disrupted and more agitated even though the solution of order $\varepsilon^0$ still applies for most of the wall layer cycle since the sweep phase is the longest.

The time scale of the wall layer is given by the Stokes solution as

$$\delta_v^+ = 3.78 t_v^+ \tag{41}$$

Where

$$t_v^+ = u_* \sqrt{\frac{t_v}{\nu}} \tag{42}$$

Since $\delta_v^+$ is essentially constant in the turbulent regime at high Reynolds numbers, so is $t_v^+$ as shown by the measurements of Meek and Baer (1971) then $t_v$ decreases as Re increases because $u_*$ increases.

In fully turbulent flow the instantaneous velocity profile has a clear inflexion point during the burst phase as shown for example by the measurements of Kim et al. (1971) in Figure 9. Following the work of Lord Rayleigh (1880) this results in a roll up of the fluid that rushes in to occupy the space vacated by the ejections. The resulting travelling vortex immediately impresses a finite fluctuation on the low speed streak in the sweep phase.

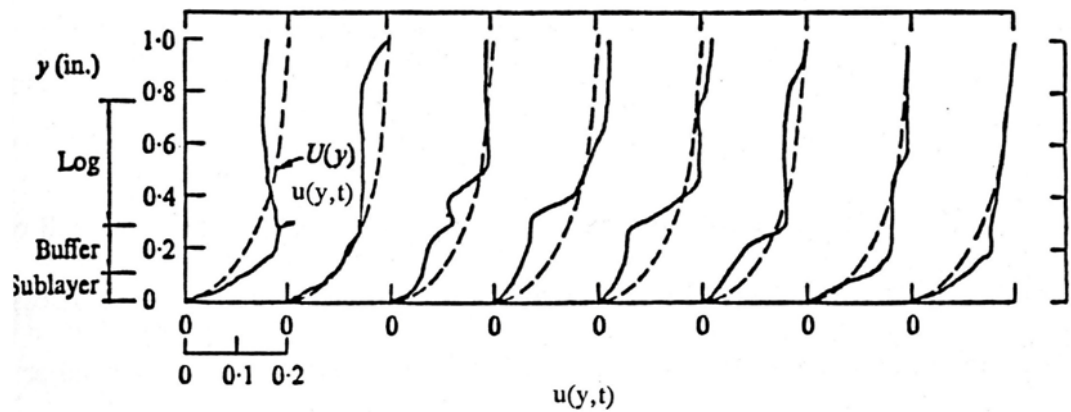

Figure 9 Smoothed phase velocity in a bursting cycle according to Kim et al. (1971)

By contrast the disturbances that come in from the pipe entrance or the leading edge of a flat plate can be weak two dimensional waves typified by Tollmien-Schlichting waves and can only grow above the minimum *critical stability Reynolds number* predicted by Tollmien and Schlichting and verified by Schubauer and Skramstadt as discussed in section 1. There is another way to visualise this critical Reynolds number.

The log-law may be viewed as a direct consequence of the ejections from the wall. In particular the Karman universal constant $\kappa$ in equation (40) can be interpreted as the slope of the shear layers associated with the ejections (Trinh, 2009c, Trinh, 2010c). As noted in section 2.2 these ejections, which are a defining element of turbulent flow, arise from the streaming flow $\psi_{st}$ part of the solution of order $\varepsilon$. The intersection between the solutions of order $\varepsilon_0$ and $\varepsilon$ represents therefore the critical point where the periodic fluctuations grow to a level where the solution of order $\varepsilon$ can no longer be neglected.

In boundary layer flow past flat plates the turbulent friction factor given by the log-law has been approximated by Prandtl as (Schlichting, 1960)

$$f = \frac{0.029}{\mathrm{Re}_x^{0.2}} \tag{43}$$

Its intersection point with the Blasius solution for laminar boundary layer flow

$$f = \frac{0.0664}{\mathrm{Re}_x^{0.5}} \tag{44}$$

occurs at Reynolds number $\mathrm{Re}_x = 34{,}000$ or $\mathrm{Re}_{\delta_1,s} = \delta_1 U_\infty / \nu = 330$ which is of the same order of magnitude as the result in equation (7).

For pipe flow, the intersection between the Hagen Poiseuille and the Prandtl-Nikuradse correlations

$$f = \frac{16}{\mathrm{Re}} \tag{45}$$

$$\frac{1}{\sqrt{f}} = 4.0 \log \mathrm{Re} \sqrt{f} - 0.4 \tag{46}$$

occurs at $Re_s = 1040$ which is again quite close to "the lowest critical Reynolds number… $Re = 1250$…(where there) appears… a saddle node bifurcation" (Eckhart et al., 2007).

Above this critical stability Reynolds number, the magnitude of the wave disturbances grow and therefore the term $\varepsilon$ increases. This means that the streaming function $\psi_{st}$ also increases. The interaction of the main flow with the streaming flow creates a three dimensional flow field, even if the original boundary layer flow is two dimensional. This interaction can modify substantially the velocity profile of the Stokes solution but the effect only becomes noticeable when a wake begins to appear behind the streaming flow. While it is not possible to predict the point where this effect occurs in wall shear flow, a similar situation can be analysed quite readily. It is well-known that surface roughness elements of scale $e$ can disturb the wall layer profile of turbulent flows ((Nikuradse, 1933) but only when the normalised roughness scale exceeds a critical value $e^+ = 5$. Nikuradse explains this limit by arguing that only roughness elements that protrude from the so-called Prandtl laminar sub-layer are effective. The modern view of the wall layer process (Figure 2) does not support Prandtl's assumption that a steady-state laminar sub-layer exists but it is easy to show that the time-averaged Stokes solution does give the relation $U^+ = y^+$ for $0 < y^+ < 4.5$ as observed by Prandtl and Nikuradse. An alternative explanation is based on the development of a wake behind a sphere (Trinh 2009) that occurs at a particle Reynolds number of $Re_p = eu/\nu = 20$ (Garner and Keey, 1958) which gives a critical roughness $e^+ = 4.5$.

When the energy contained in the streaming flow is sufficiently high it induces complete ejection of the wall fluid under the low-speed streaks into the outer region. This defines, in my view, the critical transition Reynolds number which will depend on the value of the factor

$$\varepsilon = \frac{U_e}{L\omega} = \left(\frac{U_e L}{\nu}\right)\left(\frac{\nu}{L^2 \omega}\right) \qquad (47)$$

This parameter is, in my view, a more appropriate criterion than the Reynolds number. It still contains the Reynolds number when one equates the typical length L with the diameter D but states that transition is also dependent on the frequency of the original periodic disturbances.

Thus the most appropriate critical transition parameter should be $\varepsilon_c$. The fact that there is reasonable agreement between authors on the critical transition Reynolds number $Re_{c,T}$ is because the pipe entrance in most experimental works is the same design. As noted before, Ekman (1910) and Pfenniger (1961) delayed turbulence significantly by minimising entrance disturbances. On the other hand Tam and Ghajar (1994) hastened the transition by using squared edged entrances (Figure 10). Kucur and Uzal (2007) showed that $Re_{c,T}$ is dependent on the normalised height of trip rings included in the pipe entrance.

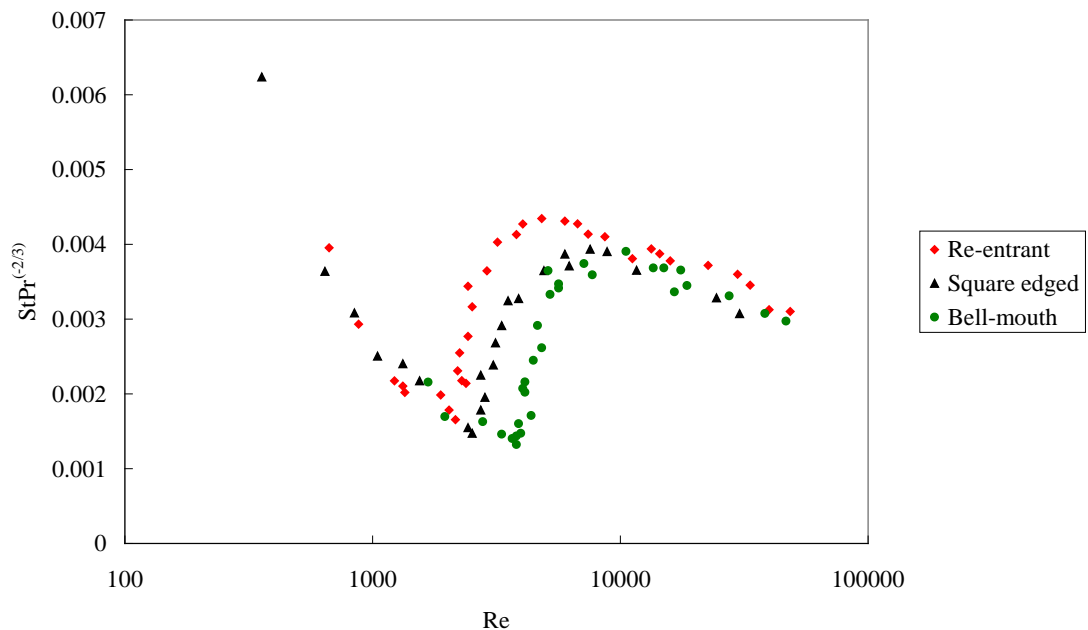

Figure 10 Effect of pipe entrance design on the critical transition Reynolds number. Data of Tam and Ghajar (1994).

The difficulty in the prediction of the critical transition Reynolds lies in estimating the position and time where the kinetic energy contained in the terms $\overline{u'_i u'_j}$ have overcome the viscous terms to eject the wall fluid. One way to address this problem is to observe that the normalised distance $y^+ = y u_* / \nu$ is a kind of local instantaneous Reynolds since it gives a ratio of kinetic to viscous forces. The end of the laminar regime occurs, according to the representation in Figure 7, when the thickness of the wall layer $\delta_v^+$ reaches an asymptotic value. For laminar boundary layer flow on a flat plate

$$\delta_v^+ = \frac{\delta_v u_*}{\nu} = \frac{\delta_v U_\infty}{\nu}\sqrt{\frac{f}{2}} = 5\,\text{Re}_x^{1/2}\left(\frac{0.664}{2\,\text{Re}_x^{1/2}}\right)^{1/2} = 2.88\,\text{Re}_x^{1/4} \qquad (48)$$

At the critical transition Reynolds number of $\text{Re}_{x,c} = 2.5 - 3\times 10^5$, this value is $\delta_v^+ = 64.5 - 67.5$ and represents the maximum penetration of viscous momentum from the wall. For pipe flow, the thickness layer of order $\varepsilon^0$ is

$$R^+ = \frac{Ru_*}{\nu} = \frac{\text{Re}}{2}\sqrt{\frac{f}{2}} = \sqrt{2\,\text{Re}} \qquad (49)$$

For $\text{Re}_{c,T} = 2100 - 2300$, $R_c^+ = 65 - 67$. In fact this critical thickness of the layer order $\varepsilon^0$ is the same for flow between parallel plates and between concentric cylinders. It appears that it is independent of the flow geometry.

In my view, the distinctive feature of turbulent flow results from the growth of the solution of order $\varepsilon$ in the region $y^+ > 65 - 67$ which is only possible when the fluid in the layer of order $\varepsilon^0$ is ejected into the region outside the wall layer. This event necessarily interrupts the further growth of the solution of order $\varepsilon^0$ which starts again with an inrush of fluid from the main flow to replenish the wall layer. As far as I know all the evidence shows that the streaming flow is not evenly distributed throughout the surface but highly localised in space and time as shown in Figure 4. Thus the inrushes that set up the low speed streaks are also localised and do not affect significantly the flow pattern in the adjacent regions. This is compatible with the formation of alternate high- and low-speed streaks first identified by Runstadler et.al. (1967). Since the normalised time scale $t_v^+$ of the solution of order $\varepsilon^0$ reaches an asymptotic value (Meek and Baer, 1970);(Trinh, 2009c), the physical time scale

$$t_v = \nu\left(\frac{t_v^+}{u_*}\right)^2 \qquad (50)$$

decreases as the Reynolds number increases. Thus the frequency of the streaming flow increases with the Reynolds number. The phenomena described here have been can be visualised in the following schematic diagram (Figure 11). In section A, the Reynolds number, the time $t_v$ required for disturbances of any strength to grow to the point of ejection is infinite.

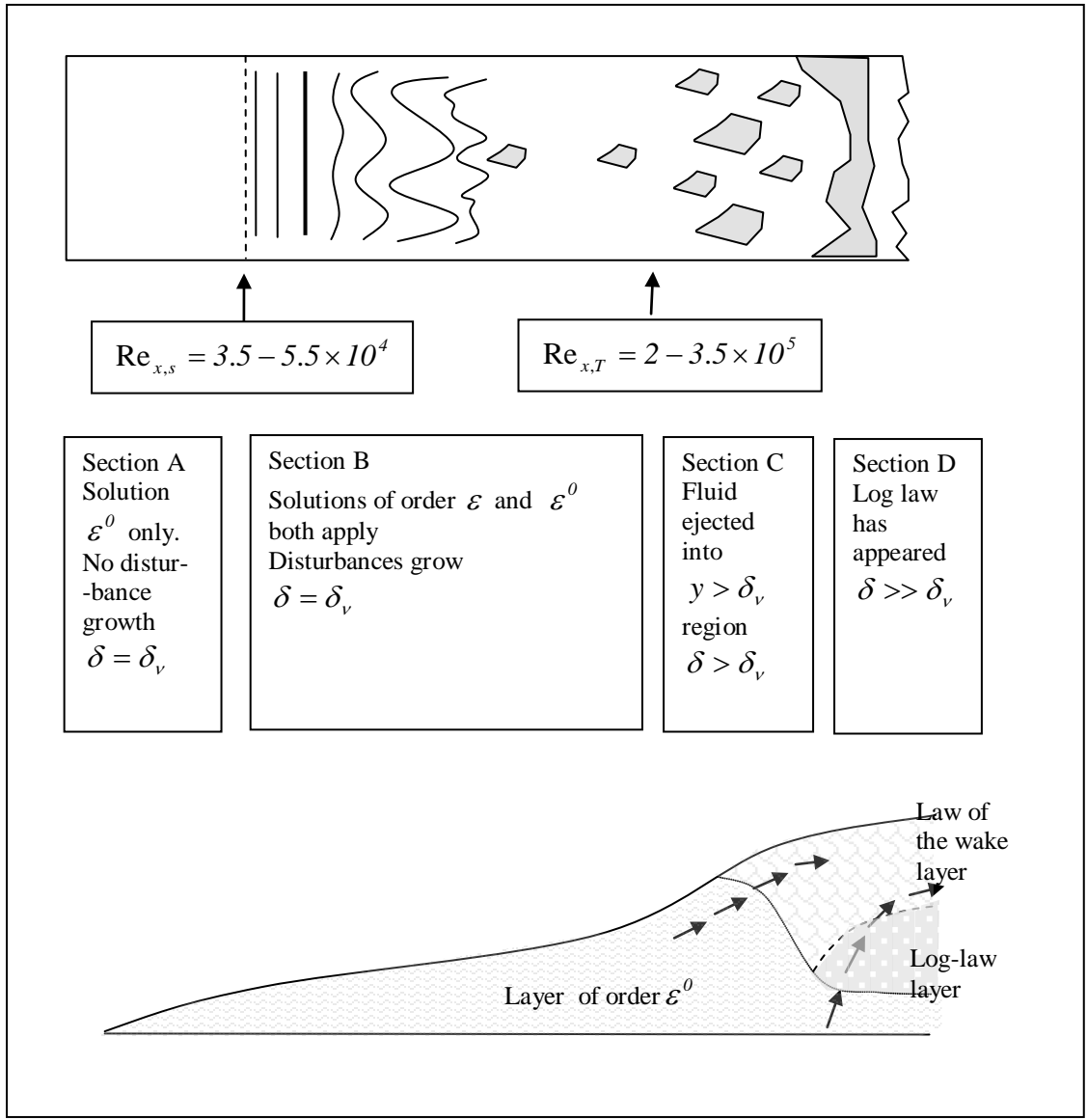

Figure 11 Transition of flow regimes along a flat plate

Therefore $\varepsilon$ is negligible and only the solution of order $\varepsilon^0$ applies: the momentum boundary layer $\delta$ is equal to the thickness $\delta_v$ of the layer of order $\varepsilon^0$. In section B, $\text{Re}_x > \text{Re}_{x,s}$ and the disturbances begin to grow the streaming flow is still weak and short lived except in the vicinity of $\text{Re}_{x,T}$ where they can disturb the solution of order $\varepsilon^0$. The data of Rothfus and Senecal (Senecal and Rothfus, 1953) in Figure 12 illustrates that the velocity distribution in pipe flow can depart from the parabolic profile predicted by the Hagen Poiseuille analysis at Re as low as *1500* but the friction factor, which is a less sensitive measure, hardly departs from equation (45).

Thus we can say that in section B, $\delta \approx \delta_v$ even though both the solutions of order $\varepsilon^0$ and $\varepsilon$ apply. The evidence comes from the proof that the time averaged Stokes solution adequately correlates all known measurements of velocity profile in the sweep phase of the wall layer (Trinh, 2010d, Einstein and Li, 1956, Meek and Baer, 1970). In addition we can transform the Stokes solution into the Blasius equation through an application of Taylor's hypothesis (Trinh and Keey, 1992b, Trinh and Keey, 1992a, Trinh, 2010b). The Blasius solution is well known to apply well in the range $\text{Re}_{x,s} < \text{Re}_x < \text{Re}_{x,T}$.

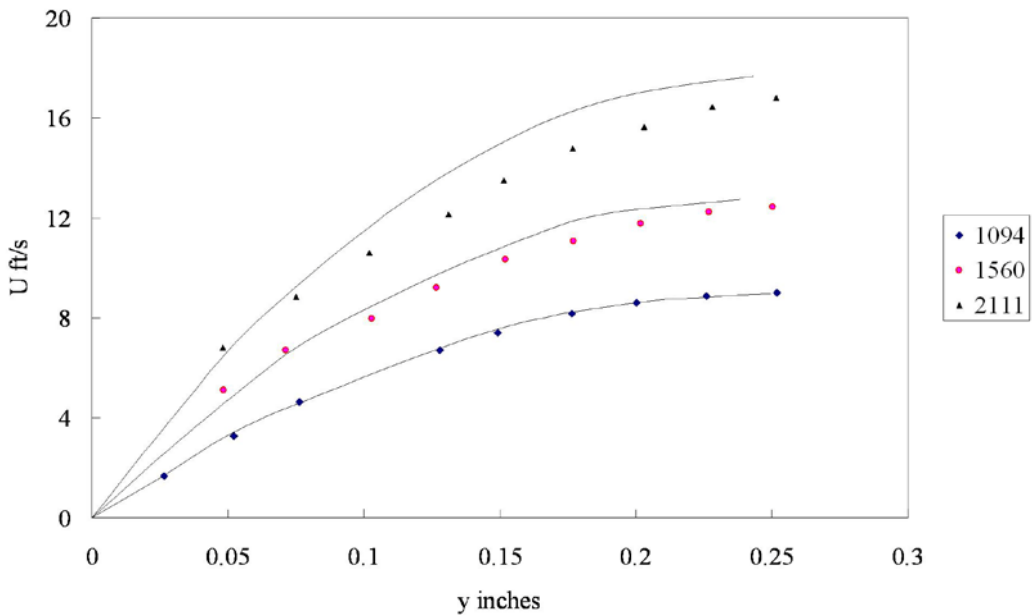

Figure 12 Velocity profile in pipe flow. Re=1500. Data of Senecal and Rothfus (1953).

In section C, the streaming flow has gathered just enough energy to eject wall fluid into the region $y^+ > \delta_v^+$. The ejected fluid is immediately deflected in the streamwise direction and forms in this visualisation a puff or slug. Immediately after $\text{Re}_{x,T}$ the time scale $t_v$ is still relatively long but the event of ejection triggers an inrush and a vortex travelling along the wall. The wall process in Figure 2 quickens rapidly with the Reynolds number because the interaction of the streamwise flow with the streaming flow requires significant energy. The time scale $t_v$ decreases exponentially with increasing Reynolds number. A parallel observation is the exponential increase of the time required for the disturbances to decay with increasing Reynolds number (Peixinho and Mullin, 2006).

In section D, the emergence of the logarithmic law of the wall, linked with the emergence of the rectilinear portion of the jet path signals the creation of wake regions where the main stream is broken up and the areas of disturbance merge into a seemingly chaotic mass.

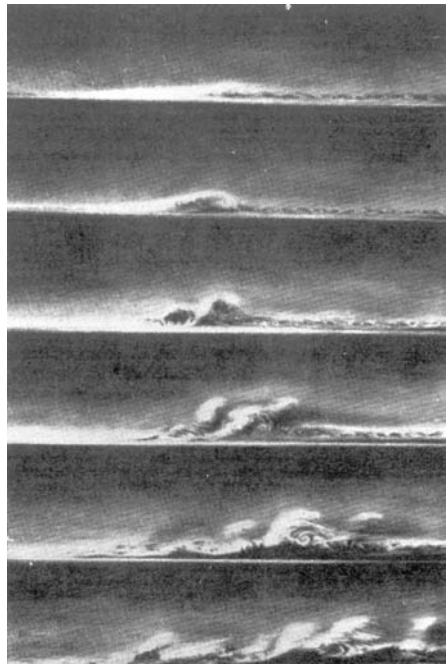

Figure 13  Flow along a flat plate; turbulence originating from a disturbance of long wavelength taken by Prandtl (1933) reproduced from (Schlichting, 1960)

The strengthening and increased frequency of the streaming flow with Reynolds number is shown clearly in the series of photos that Prandtl has taken by following an original disturbance with a camera travelling along the plate (Figure 13).

## 3  The critical transition Reynolds number in non-Newtonian fluids

It is useful to examine the transition process in non-Newtonian and drag-reducing fluids as the extra dimension allows to further probe different aspects of the problem.

## 3.1 Flat plate

The ratio of kinetic energy contained in the streaming flow to viscous resistance must be estimated at the local point of ejection. Unfortunately traditional Reynolds numbers in non-Newtonian fluid flow are based on the averaged wall shear stress. When the wall layer thickness, the friction factor and the Reynolds number are expressed in terms of the local instantaneous shear stress at the end of the sweep phase, the data for non-Newtonian fluids collapse completely onto the Newtonian curves (Trinh, 2009b). Thus the critical friction factor and Reynolds number are the same for Newtonian and non-Newtonian fluids when expressed in terms of the instantaneous shear stress. It was shown that for a flat plate the instantaneous friction factor was related to its instantaneous counterpart by

$$f = (n+1)f_e \qquad (51)$$

Then

$$\frac{f_{c,power\ law}}{f_{c,Newtonian}} = \frac{n+1}{2} \qquad (52)$$

## 3.2 Pipe flow

For pipe flow is the radius of curvature cannot be neglected and a more relevant relation is

$$\frac{f_{c,non-Newtonian}}{f_{c,Newtonian}} = \frac{3n'+1}{4n'} \qquad (53)$$

The term $(3n'+1/4n')$ appears in the Mooney-Rabinowitsch solution for laminar pipe flow (Skelland, 1967) as

$$\frac{3n'+1}{4n'} = \frac{\dot{\gamma}_w}{8V/D} = \frac{Non-Newtonian\ wall\ shear\ rate}{Newtonian\ wall\ shear\ rate} \qquad (54)$$

Its use in equation (53) is suggested by the fact that the laminar velocity profile appears as the limiting condition for $t \to \infty$ in studies of unsteady state laminar pipe flow such as the Szymanski (1932) solution adapted to turbulent flow (Trinh, 2009c).

Since the generalised Hagen Poiseuille solution is

$$f = \frac{16}{\text{Re}_g} \qquad (55)$$

Equation (53) results in a transition Reynolds number for non-Newtonian fluids as

$$\text{Re}_{g,c} = 2100 \left( \frac{3n'+1}{4n'} \right) \qquad (56)$$

There are very few experimental studies dedicated to the transition process in non-Newtonian fluids although interest is growing e.g. (Draad et al., 1998, Rudman et al., 2002, Malasova et al., 2006, Mullin and Peixinho, 2006, Peixinho et al., 2005). Most authors determine experimental values of $\text{Re}_c$ from published measurements of friction loses. We can use either a plot of wall shear stress $\tau_w$ against the flow function $(8V/D)$ (Ryan and Johnson, 1959) or the friction factor vs. Reynolds number as shown in Figure 14. The critical Reynolds number is often taken as the intersection of line (1) representing laminar flow and line (2) representing fully turbulent flow but this underestimates $\text{Re}_c$. A better estimate is given by the intersection of line (3) representing transition flow and line (2).

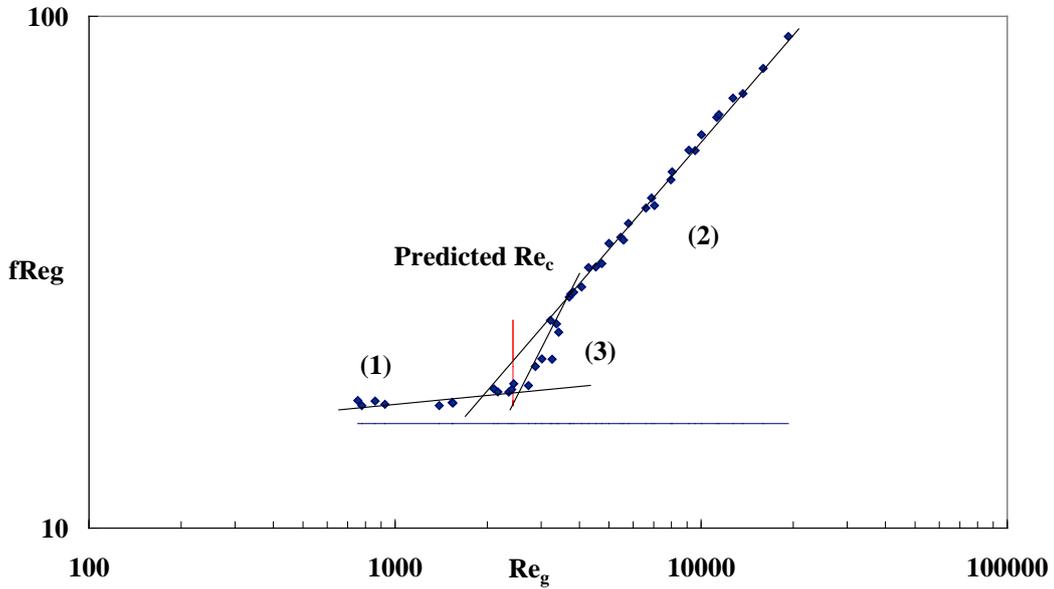

Figure 14. Estimate of $\text{Re}_{g,c}$. Data of Dodge(Dodge, 1959). 3% Carbopol, $n = .617$
(1) laminar regime (2) turbulent (3) transition.

Unfortunately transition flow varies with the entrance conditions as discussed in section 2. Many published works do not even include data points in that region.

There are also problems arising from difficulties in making accurate and relevant value of the behaviour index. When the log-log plot of $\tau_w$ vs. $(8V/D)$ is not straight line we can follow the generalisation introduced by Dodge and Metzner (Dodge and Metzner, 1959) and use the local gradient $n'$ at the point of transition. This should be based on the local instantaneous shear stress at the point of ejection as discussed in section 2.3 but only the time averaged shear stress is available in published experimental works. The difficulties in obtaining accurate measurements in non-Newtonian are evidenced in Figure 14 where the laminar flow data should follow the line $f\,\text{Re}_g = 16$ but is actually a sloping line some 10% above the exact theory. It is worth pointing out that the paper of Dodge and Metzner has been widely quoted in all subsequent papers dealing with turbulent non-Newtonian flows and most books dealing with non-Newtonian flow e.g. (Skelland, 1967, Steffe, 1996; Chhabra and Richardson, 1999). Given these uncertainties proofs of theoretical predictions of $\text{Re}_c$, including this one, should be viewed with some reservation pending the availability of more definitive and dedicated measurements. Nonetheless I have re-evaluated $\text{Re}_c$ from the data of (Dodge, 1959); (Thomas, 1960); (Caldwell and Babitt, 1941); (Carthew et al., 1983); (Güzel et al., 2009); (Malin, 1997).

Only the data of Dodge uses power law fluids and gives values of $n'$ experimentally. The data from most other researchers are based on muds and suspensions better described by the Bingham Plastic or Herschel-Bulkley models. Their laminar flow data must be reverse engineered to give an estimate of $n'$ at the point of transition. In comparing the transition in fluids obeying different rheological models, it is more useful to plot the critical frictional $f_c$ against $n'$ than $\text{Re}_c$ against $n'$. The predicted curves of Hanks, Mishra and Tripathi and equation (53) in Figure15 agree substantially in the range $0.8 < n' < 1$ but then Hanks' curve shows an exponential increase with decreasing $n'$, which is not observed in any real experimental measurement. Equation (52) does not agree with any of these curves in the range $0.3 < n' < 1$ but is closest to the curve of Mishra and Tripathi.

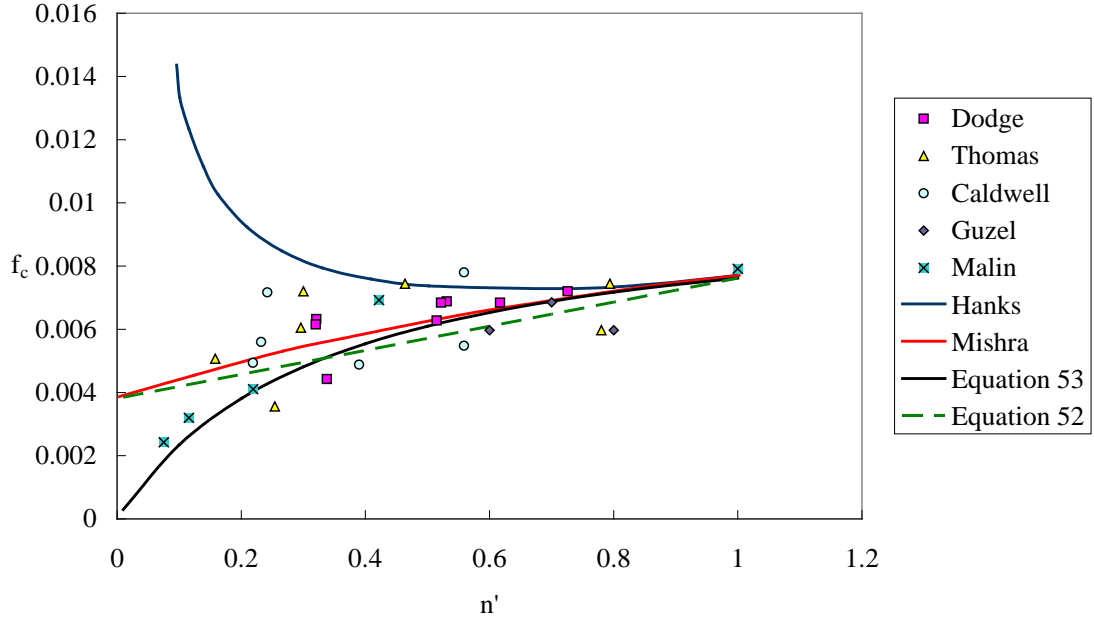

Figure 15. Critical transition friction factor vs. behaviour index $n'$. Data from Dodge (1959), Thomas (1960), Caldwell and Babbitt (1941), Güzel et al (2009), Carthew et a. (1983), Malin (1997).

Predictions from equation (53) agree with those of Mishra and Tripathi down to $n' = 0.5$ but the predictions of Mishra and Tripathi (Mishra and Tripathi, 1971) tend towards a limiting critical friction factor $f_c \approx 0.4$ while equation (53) tends asymptotically to $f_c \to 0.4$ as $n' \to 0$ indicating that the flow tends to remain laminar with increasing levels of non-Newtonian behaviour. The Mooney-Rabinowitsch equation gives (Skelland, 1967)

$$\tau_w = K'\left(\frac{8V}{D}\right)^{n'} \tag{57}$$

$$\dot{\gamma}_w = \frac{8V}{D}\left(\frac{3n'+1}{4n'}\right) \tag{58}$$

The apparent viscosity at the wall is then

$$\mu_w = K' \frac{\left(\frac{8V}{D}\right)^{n'-1}}{\left(\frac{3n'+1}{4n'}\right)} \tag{59}$$

and

$$\mu_w \to \infty \text{ as } n' \to 0 \qquad (60)$$

Therefore we should see $f \to 0$.

A different situation is encountered in drag-reducing flow. The viscoelasticity of polymer solutions slows down the growth of the perturbations and results in thicker wall layer at the point of transition (Trinh, 2010e, Trinh, 2009c). Similarly, wall riblets constrain the lateral oscillations of the low-speed streaks and increase the time scale $t_v$. As a result, the laminar flow regime is extended. Consequently, the thickness of the wall layer is increased but the wall layer velocity profile still obeys the Stokes solution (Trinh, 2009c, 2010f, 2010e). It is easy to show the relationship between the between the increased thickness of the wall layer and the drag reduction in viscoelastic fluids. For example using the velocity data of Pinho and Whitelaw (1990) to determine $\delta_v^+$ and $U_v^+$ and forcing equation (40) together with their $U_m/V$ data, we can predict the friction factor which is compared with their measurements in Figure 16.

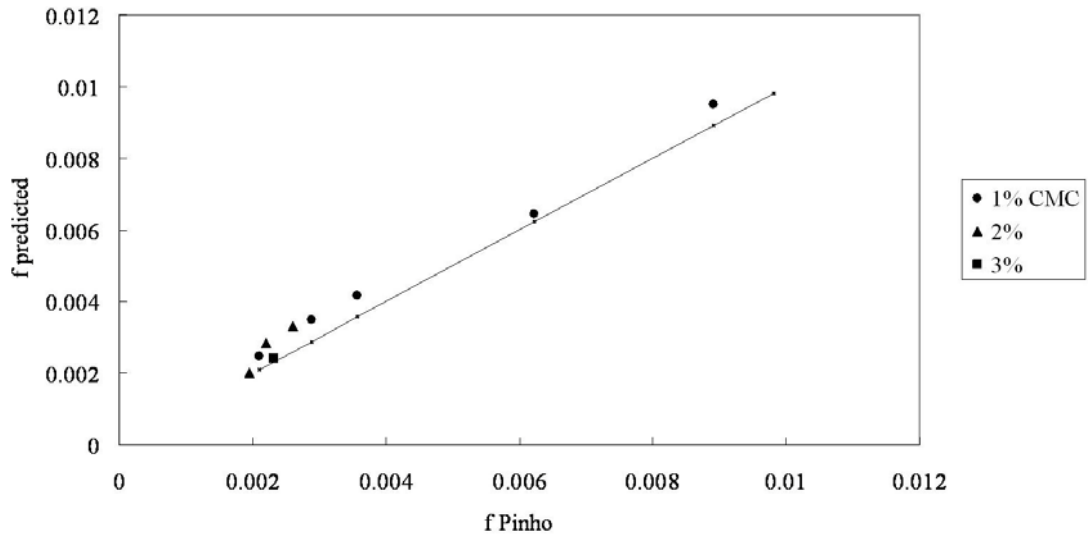

Figure 16. Comparison of calculated and experimental viscoelastic friction factors. Data of Pinho and Whitelaw (1990)

The agreement is quite reasonable considering the considerable difficulties in the measurement of accurate velocity profiles in viscoelastic flows. Some time ago (Trinh, 1969) I proposed that the effect of viscoelasticity could be approached through an analogy of the damping of waves in elastic media. Let $\Delta y$ be the shift increase in penetration of viscous diffusion. From dimensional consideration, the only parameter pertinent to the wall layer that can combined with this distance shift $\Delta y$ to give a time scale is the friction velocity $u_*$. Equation this time scale to the elastic time scale of the fluid gives

$$\left[\frac{u_*}{\Delta y}\right] = \left[\frac{\mu_a}{G}\right] \quad (61)$$

$$\left[\frac{\Delta y u_* \rho}{\mu_a}\right] = \left[\frac{\tau_w}{G}\right] \quad (62)$$

Where the ratio $\tau_w/G$ is called the Weissenberg number and represent a shift in dimensionless distance of $\delta_v^+$ above the non-elastic estimate. To estimate the value of $\text{Re}_{c,T}$ we need to know how $\tau_w/G$ varies with the shear rate $\dot{\gamma}_w$. This information is unfortunately rarely available in experimental studies of transition in viscoelastic fluids.

# 4    Conclusion

In this visualisation, the laminar boundary layer and the wall layer in turbulent flow are both associated with the solution of order $\varepsilon^0$ for flow with periodic disturbances of characteristic parameter $\varepsilon$. The solution of order $\varepsilon$ induces streaming flows that eventually lead to ejections from the wall. Transition begins when the ejections penetrate the region outside the layer of order $\varepsilon^0$. At the point of transition the thickness of the layer of order $\varepsilon^0$, $\delta_v^+$, is equal to the normalised pipe radius $R^+$. This approach allows good predictions of the critical transition friction factor in non-Newtonian pipe flow.

# 5    References

AKHAVAN, R., KAMM, R. D. & SAPHIRO, A. H. 1991. An investigation of transition to turbulence in bounded oscillatory Stokes flows. Part 2. Numerical simulations. *Journal of Fluid Mechanics,* 225**,** 423.
ANDRADE, E. N. 1931. On the circulation caused by the vibration of air in a tube. *Proc. Roy. Soc. London,* A134**,** 447.

ANTONIA, R. A., BISSET, D. K. & BROWNE, L. W. B. 1990. Effect of Reynolds number on the topology of the organized motion in a turbulent boundary layer. *Journal of Fluid Mechanics,* 213**,** 267-286.
BAYLY, B. J., ORSZAG, A. A. & HERBERT, T. 1988. Instability mechanisms in shear flow transition. *Annual Review of Fluid Mechanics,* 20**,** 359.
BIRD, R. B., STEWART, W. E. & LIGHTFOOT, E. N. 1960. *Transport Phenomena,* New York, Wiley and Sons,.
CALDWELL, D. H. & BABITT, H. E. 1941. Fow of muds, sludges and suspensions in circular pipe. *Trans, A.I.Ch.E.J.,* 25**,** 237.
CARRIERE, Z. 1929. *J. Phys. Radium,* 10**,** 198.
CARTHEW, A., GOEHRING, C. A. & VAN TEYLINGEN, J. E. 1983. Development of dynamic head loss criteria for raw sludge pumping. *Journal WPCF,* 55**,** 472-483.
CHHABRA, R. P. & RICHARDSON, J. F. 1999. *Non-Newrtonian flow in the process industries,* Oxford, Great Britain, Butterworth-Heinemann,.
COLES, D. E. 1956. The law of the wake in the turbulent boundary layer. *Journal of Fluid Mechanics,* 1**,** 191.
CORINO, E. R. & BRODKEY, R. S. 1969. A Visual investigation of the wall region in turbulent flow. *Journal of Fluid Mechanics,* 37**,** 1.
DODGE, D. W. 1959. *Turbulent flow of non-Newtonian fluids in smooth round tubes, Ph.D. Thesis,* US, University of Delaware.
DODGE, D. W. & METZNER, A. B. 1959. Turbulent Flow of Non-Newtonian Systems. *AICHE Journal,* 5**,** 189-204.
DRAAD, A. A., KUIKEN, G. D. C. & NIEUWSTADT, F. T. M. 1998. Laminar-turbulent transition in pipe flow for Newtonian and non-Newtonian fluids. *Journal Of Fluid Mechanics,* 377**,** 267-312.
DVORAK, V. 1874. *Ann. Phys. Lpz.,* 151**,** 634.
ECKHARDT, SCHNEIDER, T. M., HOF, B. & WESTERWEEL, J. 2007. Turbulence Transition in Pipe Flow. *Annu. Rev. Fluid Mech.,* 39**,** 447–68.
EINSTEIN, H. A. & LI, H. 1956. The viscous sublayer along a smooth boundary. *J. Eng. Mech. Div. ASCE 82(EM2) Pap. No 945.*
EKMAN, V. W. 1910. On the change from steady to turbulent motion of liquids. *Ark. f. Mat. Astron. och Fys.,* 6.
ETEMAD, S. G. & SADEGHI, M. 2001. Non-Newtonian pressure drop and critical Reynolds number through rectangular duct. *International Communications In Heat And Mass Transfer,* 28**,** 555-563.
FARADAY, M. 1831. On a Peculiar Class of Acoustical Figures; and on Certain Forms Assumed by Groups of Particles upon Vibrating Elastic Surfaces. *Phil. Trans. R. Soc. Lond. ,* 121**,** 299-340.
GAD-EL-HAK, M., DAVIS, S. H., MCMURRAY, J. T. & ORSZAG, S. A. 1983. On the stability of the decelerating laminar boundary layer. *Journal of Fluid Mechanics***,** 297-323.
GARNER, F. H. & KEEY, R. B. 1958. Mass-transfer from single solid spheres—I : Transfer at low Reynolds numbers *Chem. Eng. Sci,* 9.
GÜZEL, B., FRIGAARD, I. & MARTINEZ, D. M. 2009. Predicting laminar–turbulent- transition in Poiseulle pipe flow for non-Newtonian fluids. *Chemical EngineeringScience,* 64**,** 254--264.
HANKS, R. W. 1963a. The laminar-turbulent transition for fluids with a yield stress. *AIChE J.,* 9**,** 306-309.

HANKS, R. W. 1963b. The laminar-turbulent transition in pipes, concentric annuli, and parallel plates. *AIChE J.,* 9**,** 45-48.
HANKS, R. W. & DADIA, B. H. 1971. Theoretical Analysis of Turbulent Flow of Non-Newtonian Slurries in Pipes. *AICHE Journal,* 17**,** 554-&.
HANKS, R. W. & RICKS, B. L. 1975. Transitional and turbulent pipe flow of pseudoplastic fluids. *Journal of Hydronautics,* 9**,** 39-44.
HANKS, R. W. & VALIA, M. P. 1971. Theory of Transitional and Turbulent Flow of Non-Newtonian Slurries between Flat Parallel Plates. *Society of Petroleum Engineers Journal,* 11**,** 52-&.
HANRATTY, T. J. 1956. Turbulent exchange of mass and momentum with a boundary. *AIChE J.,* 2**,** 359.
HERBERT, T. 1988. Secondary instability of boundary layers. *Annual Review of Fluid Mechanics,* 20**,** 487.
HOF, B., JUEL, A. & MULLIN, T. 2003. Scaling of the turbulence transition threshold in a pipe. *Phys. Rev. Lett.,* 91**,** 244502.
KERSWELL, R. R. 2005. Recent progress in understanding the transition to turbulence in a pipe. *Nonlinearity,* 18**,** R17–R44.
KIM, H. T., KLINE, S. J. & REYNOLDS, W. C. 1971. The Production of the Wall Region in Turbulent Flow. *Journal of Fluid Mechanics,* 50**,** 133.
KLINE, S. J., REYNOLDS, W. C., SCHRAUB, F. A. & RUNSTADLER, P. W. 1967. The structure of turbulent boundary layers. *Journal of Fluid Mechanics,* 30**,** 741.
KUCUR, M. & UZAL, E. 2007. Controlling critical Reynolds number of parallel plate flow by boundary input. *Aircraft Engineering and Aerospace Technology: An International Journal,* 79**,** 507–510.
LIN, C. C. 1955. The Theory of Hydrodynamic Stability. *Cambridge University Press*.
MALASOVA, I., MALKIN, A. Y., KHARATIYAN, E. & HALDENWANG, R. 2006. Scaling in pipeline flow of Kaolin suspensions. *J. Non-Newtonian Fluid Mech.,* 136**,** 76-78.
MALIN, M. R. 1997. The Turbulent Flow of Bingham Plastic Fluids in Smooth Circular Tubes. *Int. Com. Heat Mass Transfer,* 24**,** 793-804.
MEEK, R. L. & BAER, A. D. 1970. The Periodic viscous sublayer in turbulent flow. *AIChE J.,* 16**,** 841.
MESEGUER, A. & TREFETHEN, L. N. 2003. Linearized pipe flow to Reynolds number 107. *J. Comput. Phys.,* 186**,** 178.
METZNER, A. B. & REED, J. C. 1955. Flow of Non-Newtonian Fluids - Correlation of the Laminar, Transition, and Turbulent-Flow Regions. *Aiche Journal,* 1**,** 434-440.
MILLIKAN, C. B. 1938. A critical discussion of turbulent flows in channels and circular tubes. *In:* DEN HARTOG, J. P. & PETERS, H. (eds.) *5th Int. Congr. Appl. Mech. Proc.*: Wiley.
MISHRA, P. & TRIPATHI, G. 1971. Transition from laminar to turbulent flow of purely viscous non-Newtonian fluids in tubes. *Chemical Engineering Science,* 26**,** 915-921.
MISHRA, P. & TRIPATHI, G. 1973. Heat And Momentum-Transfer To Purely Viscous Non-Newtonian Fluids Flowing Through Tubes. *Transactions Of The Institution Of Chemical Engineers,* 51**,** 141-150.
MOIN, P. & KIM, J. 1982. Numerical Investigation of Turbulent Channel Flow. *Journal of Fluid Mechanics,* 118**,** 341-377.

MULLIN, T. & PEIXINHO, J. 2006. Transition to Turbulence in Pipe Flow. *Journal of Low Temperature Physics,* 145**,** 75-88.
NIKURADSE, J. 1933. Stromungsgesetz in rauhren rohren. *VDIForschungshefte***,** 361.
OBUKHOV, A. M. 1983. Kolmogorov flow and laboratory simulation of it. *Russian Mathematical Surveys,* 38**,** 113-126.
ORR, W. M. F. 1907. The stability or instability of the steady motions of a perfect liquid and a viscous fluid. *Proc. Roy. Ir. Acad.,* A27**,** 689.
PEIXINHO, J. & MULLIN, T. 2006. Decay of Turbulence in Pipe Flow. *Physical Review Letters***,** PRL 96, 094501.
PEIXINHO, J., NOUAR, C., DESAUBRY, C. & THERON, B. 2005. Laminar transitional and turbulent flow of yield stress fluid in a pipe. *Journal of Non-Newtonian Fluid Mechanics,* 128**,** 172-184.
PFENNIGER, W. 1961. *In:* LACHMAN, G. V. (ed.) *Boundary Layer and Flow Control.* Pergamon.
PINHO, F. T. & WHITELAW, J. H. 1990. Flow Of Non-Newtonian Fluids In A Pipe. *Journal Of Non-Newtonian Fluid Mechanics,* 34**,** 129-144.
PRANDTL, L. 1933. Neuere Ergebnisse der Turbulenzforschung. . *Z.VDI,* 77**,** 105-114.
RAYLEIGH, L. 1880. On the stability of certain fluid motions. *Proc. London Math. Soc.,* 11**,** 57.
RAYLEIGH, L. 1884. On the circulation of air observed in Kundt's tubes and on some allied accoustical problems. *Phil.Trans.Roy.Soc.Lond,* 175**,** 1-21.
REYNOLDS, O. 1883. An experimental investigation of the circumstances which determine whether the motion of water shall be direct or sinous, and of the law of resistances in parallel channels. *Phil Trans Roy Soc London,* 174**,** 935-982.
REYNOLDS, O. 1895. On the Dynamical theory of Incompressible Viscous Fluids and the Determination of the Criterion. *Phil. Trans. Roy. Soc. (London),* 186A**,** 123-164.
RILEY, N. 1967. *J. Inst. Math. Appl.,* 3**,** 419.
ROBINSON, S. K. 1991. Coherent Motions in the Turbulent Boundary Layer. *Annual Review of Fluid Mechanics,* 23**,** 601.
RUDMAN, M., GRAHAM, L. J. H., BLACKBURN, M. & PULLUM, L. 2002. Non-Newtonian Turbulent and Transitional Pipe Flow. *Hy15* Banff, Canada.
RYAN, N. W. & JOHNSON, M. M. 1959. Transition from laminar to turbulent flows in pipes. *AIChE J.,* 5.
SALWEN, H., COTTEN, F. W. & GROSCH, C. E. 1980. Linear stability of Poiseuille flow in a circular pipe. *J. Fluid Mech.,* 98**,** 273.
SCHLICHTING, H. 1932. Berechnung ebener periodischer Grenzschichtstromungen. *Phys.Z.,* 33**,** 327-335.
SCHLICHTING, H. 1960. *Boundary Layer Theory,* New York., MacGrawHill.
SCHLICHTING, H. 1979. *Boundary layer theory. Seventh edition.*, McGraw-Hill.
SCHNECK, D. J. & WALBURN, F. J. 1976. Pulsatile blood flow in a channel of small exponential divergence Part II: Steady streaming due to the interaction of viscous effects with convected inertia. *Journal of Fluids Engineering***,** 707.
SCHUBAUER, G. B. & SKRAMSTAD, H. K. 1943. Laminar boundary layer oscillations and transition on a flat plate. NACA, Rep. No 909

SENECAL, V. E. & ROTHFUS, R. R. 1953. Transition flow of fluids in smooth tubes. *Chem. Eng. Prog.,* 49**,** 533.

SKELLAND, A. H. P. 1967. *Non-Newtonian Flow and Heat transfer,* New York, John Wiley and Sons.
SOMMERFIELD, A. 1908. Ein Beitrag zur hydrodynamischen Erklarung der turbulenten Flussigkeisbewegung. *Atti Congr. Int. Math. 4th* Rome.
STEFFE, J. F. 1996. *Rheological Methods in Food Process Engineering,* USA, Freeman Press.
STOKES, G. G. 1851. On the effect of the internal friction of fluids on the motion of pendulums. *Camb. Phil. Trans.,* IX**,** 8.
STUART, J. T. 1966. Double boundary layers in oscillatory viscous flow. *Journal of Fluid Mechanics,* 24**,** 673.
SZYMANSKI, P. 1932. Quelques solutions exactes des équations de l'hydrodynamique de fluide visqueux dans le cas d'un tube cylindrique. *J. des Mathematiques Pures et Appliquées,* 11 series 9**,** 67.
TAM, L. & GHAJAR, A. J. 1994. Heat Transfer Measurements and Correlations in the Transition Region for a Circular Tube with Three Different Inlet Configurations. *Experimental Thermal and Fluid Science,* 8**,** 79-90.
TETLIONIS, D. M. 1981. Unsteady Viscous Flow. *SpringerVerlag, New York.*
THOMAS, A. D. 1960. Heat and momentum transport characteristics of of non-Newtonian aqueous Thorium oxide suspensions. *AIChEJ,* 8.
TRINH, K. T. 1969. *A boundary layer theory for turbulent transport phenomena, M.E. Thesis,,* New Zealand, University of Canterbury.
TRINH, K. T. 1992. *Turbulent transport near the wall in Newtonian and non-Newtonian pipe flow, Ph.D. Thesis,* New Zealand, University of Canterbury.
TRINH, K. T. 2005. A zonal similarity analysis of wall-bounded turbulent shear flows. *7th World Congress of Chemical Engineering: Proceedings.* Glasgow.
TRINH, K. T. 2009a. A Four Component Decomposition of the Instantaneous Velocity in Turbulent Flow. *arXiv.org 0912.5248v1 [phys.fluid-dyn]* [Online].
TRINH, K. T. 2009b. The Instantaneous Wall Viscosity in Pipe Flow of Power Law Fluids: Case Study for a Theory of Turbulence in Time-Independent Non-Newtonian Fluids. *arXiv.org 0912.5249v1 [phys.fluid-dyn]* [Online].
TRINH, K. T. 2009c. A Theory Of Turbulence Part I: Towards Solutions Of The Navier-Stokes Equations. *arXiv.org 0910.2072v1 [physics.flu.dyn.]* [Online].
TRINH, K. T. 2010a. Additive Layers: An Alternate Classification Of Flow Regimes. *arXiv.org 1001.1587 {phys.fluid-dyn]* [Online].
TRINH, K. T. 2010b. The Fourth Partial Derivative In Transport Dynamics. *arXiv.org 1001.1580[math-ph.]* [Online].
TRINH, K. T. 2010c. On the Karman constant. *arxiv.org/abs/1007.0605 [phys.fluid-dyn]* [Online].
TRINH, K. T. 2010d. On the Non Specific Nature of Classical Turbulence Statistics. *arXiv.org 1001.2064 [phys.flu-dyn]* [Online].
TRINH, K. T. 2010e. On Virk's Asymptote. *arXiv.org 1001.1582[phys.fluid-dyn]* [Online].
TRINH, K. T. 2010f. A Zonal Similarity Analysis of Velocity Profiles in Wall-Bounded Turbulent Shear Flows. *arXiv.org 1001.1594 [phys.fluid-dyn]* [Online].
TRINH, K. T. & KEEY, R. B. 1992a. A Modified Penetration Theory and its Relation to Boundary Layer Transport. *Trans. IChemE, ser. A,* 70**,** 596-603.
TRINH, K. T. & KEEY, R. B. 1992b. A Time-Space Transformation for Non-Newtonian Laminar Boundary Layers. *Trans. IChemE, ser. A,* 70**,** 604-609.

WILSON, K. C. & THOMAS, A. D. 2006. Analytic model of laminar-turbulent transition for Bingham plastics. *Canadian Journal of Chemical Engineering,* 84**,** 520-526.
ZAMORA, M., ROY, S. & SLATER, K. 2005. Comparing a Basic Set of Drilling Fluid Pressure-Loss Relationships to Flow-Loop and Field Data. *AADE 2005 National Technical Conference and Exhibition.* Houston, Texas.